\documentclass[
aps,
prl,
longbibliography,
superscriptaddress,
floatfix,
preprint
]{revtex4-1}

\usepackage{mathrsfs}
\usepackage{amsmath}
\usepackage{amsfonts}
\usepackage{amssymb}
\usepackage{amsthm}
\usepackage{graphicx}
\usepackage{color}
\usepackage[pdfpagelayout=OnePageLeft]{hyperref}
\usepackage{bm}
\usepackage[caption=false]{subfig}
\usepackage{verbatim}
\usepackage[per-mode=symbol,separate-uncertainty]{siunitx}
\usepackage{booktabs}
\usepackage{xspace}

\DeclareSIUnit\dBm{dBm}
\DeclareSIUnit\dB{dB}
\DeclareSIUnit\inch{in}

\begin{document}
\title{Exchange-enhanced Ultrastrong Magnon-Magnon Coupling in a Compensated Ferrimagnet}

\author{Lukas~Liensberger}
\email[]{Lukas.Liensberger@wmi.badw.de}
\affiliation{Walther-Mei{\ss}ner-Institut, Bayerische Akademie der Wissenschaften, 85748 Garching, Germany}
\affiliation{Physik-Department, Technische Universit\"{a}t M\"{u}nchen, 85748 Garching, Germany}

\author{Akashdeep~Kamra}
\email[]{Akashdeep.Kamra@ntnu.no}
\affiliation{Center for Quantum Spintronics, Department of Physics, Norwegian University of Science and Technology, 7491 Trondheim, Norway}

\author{Hannes~Maier-Flaig}
\affiliation{Walther-Mei{\ss}ner-Institut, Bayerische Akademie der Wissenschaften, 85748 Garching, Germany}
\affiliation{Physik-Department, Technische Universit\"{a}t M\"{u}nchen, 85748 Garching, Germany}

\author{Stephan~Gepr\"{a}gs}
\affiliation{Walther-Mei{\ss}ner-Institut, Bayerische Akademie der Wissenschaften, 85748 Garching, Germany}

\author{Andreas~Erb}
\affiliation{Walther-Mei{\ss}ner-Institut, Bayerische Akademie der Wissenschaften, 85748 Garching, Germany}

\author{Sebastian~T.~B.~Goennenwein}
\affiliation{Institut f\"{u}r Festk\"{o}rper- und Materialphysik, Technische Universit\"{a}t Dresden, 01062 Dresden, Germany}

\author{Rudolf~Gross}
\affiliation{Walther-Mei{\ss}ner-Institut, Bayerische Akademie der Wissenschaften, 85748 Garching, Germany}
\affiliation{Physik-Department, Technische Universit\"{a}t M\"{u}nchen, 85748 Garching, Germany}
\affiliation{Nanosystems Initiative Munich, 80799 Munich, Germany}
\affiliation{Munich Center for Quantum Science and Technology (MCQST), 80799 Munich, Germany}

\author{Wolfgang~Belzig}
\affiliation{Department of Physics, University of Konstanz, 78457 Konstanz, Germany}

\author{Hans~Huebl}
\affiliation{Walther-Mei{\ss}ner-Institut, Bayerische Akademie der Wissenschaften, 85748 Garching, Germany}
\affiliation{Physik-Department, Technische Universit\"{a}t M\"{u}nchen, 85748 Garching, Germany}
\affiliation{Nanosystems Initiative Munich, 80799 Munich, Germany}
\affiliation{Munich Center for Quantum Science and Technology (MCQST), 80799 Munich, Germany}

\author{Mathias~Weiler}
\email[]{Mathias.Weiler@wmi.badw.de}
\affiliation{Walther-Mei{\ss}ner-Institut, Bayerische Akademie der Wissenschaften, 85748 Garching, Germany}
\affiliation{Physik-Department, Technische Universit\"{a}t M\"{u}nchen, 85748 Garching, Germany}

\date{\today}

\begin{abstract}
We experimentally study the spin dynamics in a gadolinium iron garnet single crystal using broadband ferromagnetic resonance. Close to the ferrimagnetic compensation temperature, we observe ultrastrong coupling of clockwise and counterclockwise magnon modes. The magnon-magnon coupling strength reaches almost 40\% of the mode frequency and can be tuned by varying the direction of the external magnetic field. We theoretically explain the observed mode-coupling as arising from the broken rotational symmetry due to a weak magnetocrystalline anisotropy. The effect of this anisotropy is exchange-enhanced around the ferrimagnetic compensation point. 
\end{abstract} 

\maketitle

The strong and ultrastrong interaction of light and matter is foundational for circuit quantum electrodynamics~\cite{FriskKockum2019, Zhu2011, Viennot2015}. The realizations of strong spin-photon~\cite{Schuster2010,Kubo2010,Samkharadze2018} and magnon-photon~\cite{Huebl2013, Zhang2014, Bai2015, Liu2016, ViolaKusminskiy2016, Harder2018} coupling have established magnetic systems as viable platforms for frequency up-conversion~\cite{Hisatomi2016, Klingler2016} and quantum state storage~\cite{Tabuchi2015}. Antiferromagnets and ferrimagnets further host multiple magnon modes. Their coupling allows for coherent control and engineering of spin dynamics for applications in magnonics~\cite{Kruglyak2010,Chumak2015} and antiferromagnetic spintronics~\cite{Jungwirth2016, Baltz2018}.

Recently, it has been shown~\cite{Klingler2018, Chen2018, Qin2018} that the weak interlayer exchange interaction between two magnetic materials can cause magnon-magnon coupling. However, the much stronger intrinsic exchange has not yet been leveraged for coupling phenomena. While the THz-frequency dynamics in antiferromagnets is challenging to address experimentally~\cite{Kampfrath2011}, the sublattice magnetizations in compensated ferrimagnets can be tuned to achieve GHz-frequency quasi-antiferromagnetic dynamics. Here, we report the experimental observation of ultrastrong exchange-enhanced magnon-magnon coupling in a compensated ferrimagnet with the coupling rate reaching up to 37\% of the characteristic magnon frequency. We furthermore demonstrate that the coupling strength can be continuously tuned from the ultrastrong to the weak regime.

\begin{figure}
	\begin{center}
		\includegraphics[]{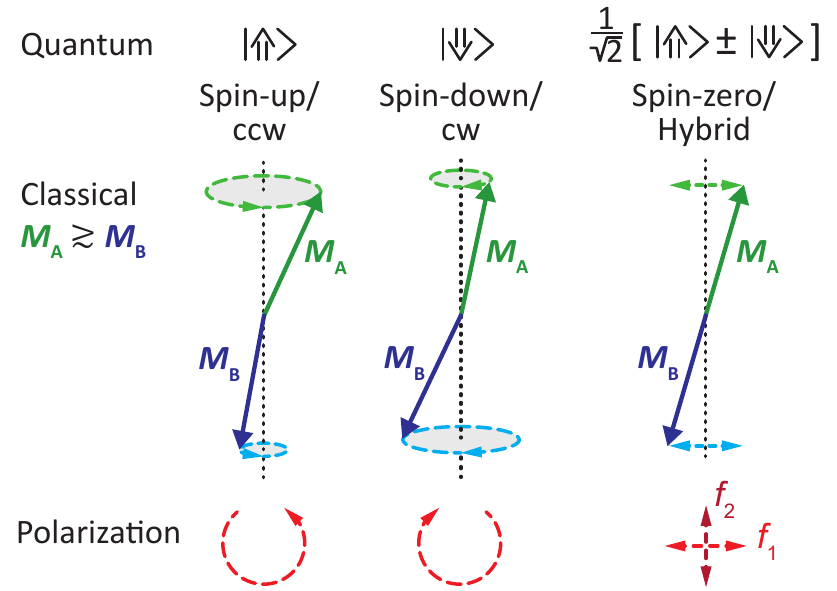}
		\caption{Classical and quantum representations of the magnetization dynamics in a two-sublattice compensated ferrimagnet. The eigenmodes of the compensated ferrimagnet close to its compensation temperature are similar to that of an antiferromagnet since the sublattice magnetizations are almost identical (we choose $M_\mathrm{A} \gtrsim M_\mathrm{B}$). In the quantum picture, the classical modes with counter-clockwise (ccw) and clockwise-precession (cw) are identified as spin-up and spin-down magnons. The hybridized modes with linear polarization correspond to spin-zero magnons~\cite{KamraPRB2017}. The angles between the two sublattice magnetizations have been exaggerated for clarity.}
		\label{fig:main}
	\end{center}
\end{figure}

We investigate spin dynamics, or equivalently the magnon modes, in a compensated, effectively two-sublattice ferrimagnet in the collinear state. Around its compensation temperature, this system can be viewed as a ``quasi-antiferromagnet'' due to its nearly identical sublattice magnetizations $M_\mathrm{A} \gtrsim M_\mathrm{B}$. Figure~\ref{fig:main} schematically depicts the typical spatially uniform spin dynamics eigenmodes of the system~\cite{gurevich1996magnetization}. 
Within the classical description, these become clockwise (cw) and counterclockwise (ccw) precessing modes which correspond to spin-down and spin-up magnons, respectively, in the quantum picture. The key physics underlying our experiments is the tunable exchange-enhanced coupling, and the concomitant hybridization, between theses two modes. The essential ingredients - mode coupling and exchange-enhancement - are both intuitively understood within the quantum picture as follows. First, due to their opposite spins, a spin-up magnon can only be coupled to its spin-down counterpart by a mechanism that violates the conservation of spin along the sublattice magnetization, and thus magnon spin, direction~\cite{KamraPRB2017}. Since angular momentum conservation is a consequence of rotational invariance or isotropy, an anisotropy about the magnon spin axis provides such a coupling mechanism. Achieving the equilibrium sublattice magnetizations, or equivalently the magnon spin axis, to lie along directions with varying degrees of local axial anisotropy allows to correspondingly vary the resultant magnon-magnon coupling. This explains the nonzero mode-coupling along with its tunability. 
However, the typically weak magnetocrystalline anisotropy may not be expected to yield observable effects and, therefore, has typically been disregarded. This is where exchange-enhancement in a quasi-antiferromagnet makes the crucial difference. The antiferromagnetic magnons, despite their unit net spin, are formed by large, nearly equal and opposite spins on the two sublattices~\cite{Kamra2019}. The anisotropy-mediated mode coupling results from, and is proportional to, this large sublattice spin instead of the unit net spin, and is therefore strongly amplified. This amplification effect is termed exchange-enhancement within the classical description~\cite{Kamra2019,Keffer1952,Kamra2018}.

\begin{figure*}
	\begin{center}
	\includegraphics[]{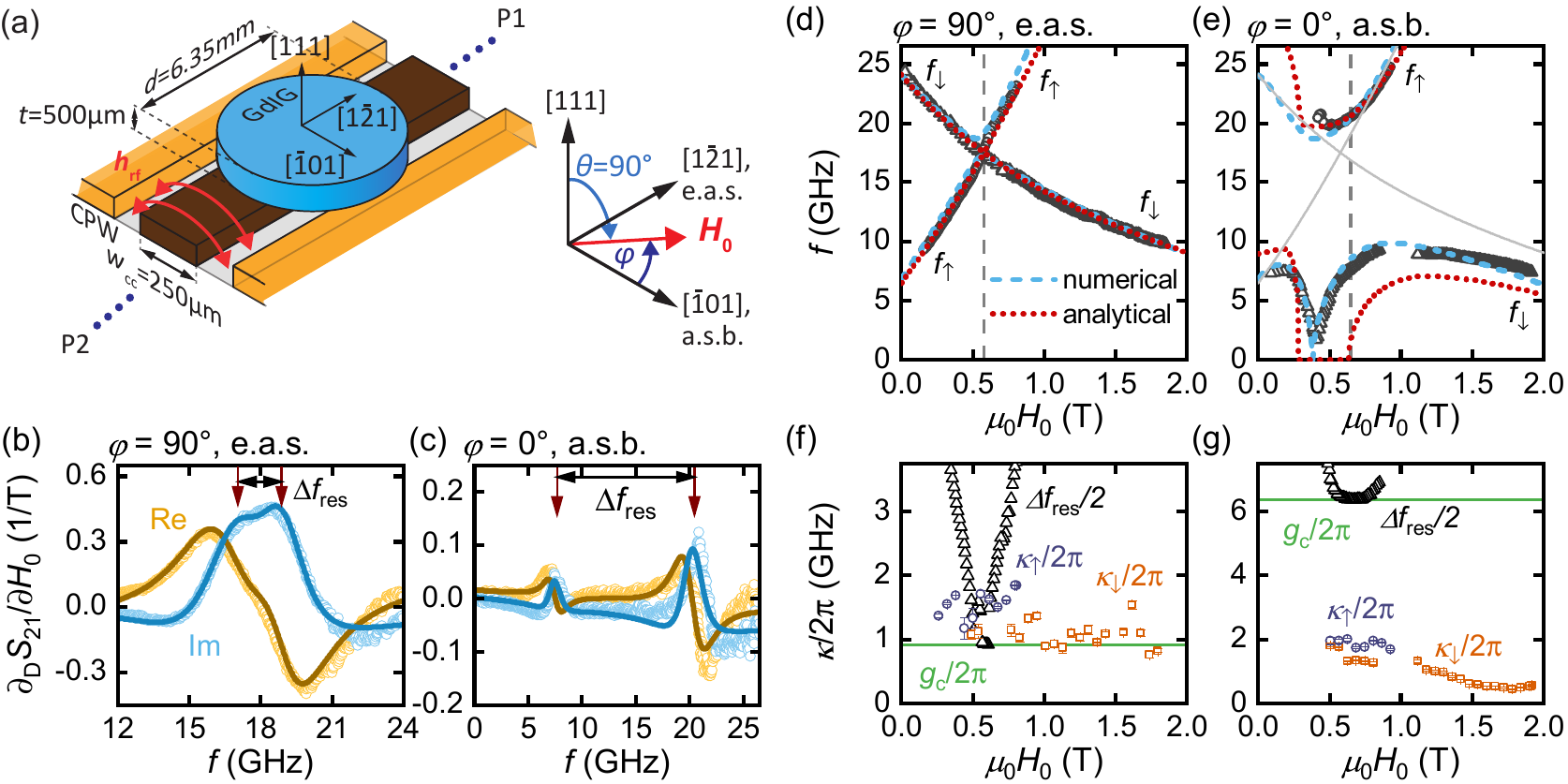}
	\caption{(a)~Schematic broadband ferromagnetic resonance (BMR) setup, with the GdIG disk on the coplanar waveguide (CPW). The angle $\varphi$ defines the in-plane direction of the magnetic field $\bm{H}_0$.
	(b),(c)~BMR spectra obtained for fixed magnetic field strengths applied along the (b) effectively axially symmetric (e.a.s.) direction in the (111)-plane at $\varphi=\ang{90}$ ($\mu_0H_0=\SI{0.58}{\tesla}$) and along the (c) axial symmetry broken (a.s.b.) axis $\varphi=\ang{0}$ ($\mu_0H_0=\SI{0.65}{\tesla}$) recorded at $T=\SI{282}{\kelvin}$ ($T_\mathrm{comp}=\SI{288}{\kelvin}$). 
	The solid lines are fits to Eq.~(S7)~\cite{SupInf}. The resonance frequencies are indicated by the red arrows and their difference is denoted as $\Delta f_\mathrm{res}$.  
	(d),(e)~Mode frequencies vs.\ applied magnetic field strength measured at $T=\SI{282}{\kelvin}$ where $M_{\mathrm{Gd}} \gtrsim M_{\mathrm{Fe}}$. Open circles and triangles denote measured resonance frequencies. The red dotted curves depict results of our analytical model and the blue dashed lines are obtained by numerical simulation. Along the e.a.s.\ direction $\varphi=\ang{90}$ (d), weak coupling is observed, whereas along the a.s.b.\ direction $\varphi=\ang{0}$ (e), we find ultrastrong coupling (see text). The solid gray lines in (e) indicate the uncoupled case taken from the analytical solution of panel (d). 
	(f),(g)~Linewidths $\kappa/2\pi$ of the spin-up $\kappa_\uparrow$ and spin-down $\kappa_\downarrow$ modes, and resonance frequency splitting $\Delta f_\mathrm{res}/2$ as a function of $H_0$. The coupling strength $g_\mathrm{c}/2\pi$ is given by the minimum of ${\Delta f_\mathrm{res}/2}$.
	}
	\label{fig:setup}
	\end{center}
\end{figure*}

In our corresponding experiments, we study the magnetization dynamics of a (111)-oriented single crystal Gd$_3$Fe$_5$O$_{12}$ (gadolinium iron garnet, GdIG) disk by broadband magnetic resonance (BMR)~\cite{Kalarickal2006}. A schematic depiction of the setup is shown in Fig.~\ref{fig:setup}(a). We use a vector network analyzer to record the complex transmission $S_{21}$ as a function of the microwave frequency $f$ and the external magnetic field $\bm{H}_0$ applied in the (111)-plane.  Our experiments are performed at $T=\SI{282}{\kelvin}$, slightly below the ferrimagnetic compensation point $T_\mathrm{comp}=\SI{288}{\kelvin}$, as determined by SQUID-magnetometry~\cite{SupInf}. At this temperature, the resonance frequencies of the spin-up and spin-down modes are in the microwave frequency range.

In Fig.~\ref{fig:setup}(b), we show the normalized background-corrected field-derivative of $S_{21}$~\cite{Maier-Flaig2018} recorded at fixed magnetic field magnitude $\mu_0H_0=\SI{0.58}{\tesla}$ applied at $\varphi=\ang{90}$. As discussed later, this is a situation in which the magneto-crystalline anisotropy energy has axial symmetry about the magnetic field direction. We refer to this case as an effectively axially symmetric (e.a.s.) direction. By fitting the data to Eq.~(S7)~\cite{SupInf}, we extract the resonance frequencies $f_1$ and $f_2$ of the two observed resonances, their difference $\Delta f_\mathrm{res}$ and their linewidths $\kappa_1$ and $\kappa_2$. In Fig.~\ref{fig:setup}(c) we show corresponding data and fits for $\varphi=\ang{0}$ and $\mu_0H_0=\SI{0.65}{\tesla}$, which corresponds to a situation in which the magneto-crystalline anisotropy energy is anisotropic about the applied magnetic field direction, which we refer to as an axial symmetry broken (a.s.b.) direction, as explained below. Again, two resonances are observed. In contrast to the data in Fig.~\ref{fig:setup}(b), the resonances are now clearly separated.

We repeat these experiments for a range of magnetic field magnitudes $H_0$ applied along the two directions (e.a.s.\ and a.s.b.) of interest. The obtained resonance frequencies are shown as symbols in Figs.~\ref{fig:setup}(d) and~(e). In the e.a.s.\ case shown in Fig.~\ref{fig:setup}(d), we clearly observe two resonance modes. The first one follows $\partial f_\mathrm{res}/\partial H_0 > 0$ and is the spin-up mode $f_\uparrow$ and the second resonance with $\partial f_\mathrm{res}/\partial H_0 < 0$ is the spin-down mode $f_\downarrow$. The vertical dashed line corresponds to $\mu_0H_0=\SI{0.58}{\tesla}$ where $\Delta f_\mathrm{res}$ is minimized and the data shown in Fig.~\ref{fig:setup}(b) is obtained. The resonance frequencies are in excellent agreement with those obtained from numerical (see Supplemental Material~\cite{SupInf}) and analytical (see below) solutions to the Landau-Lifshitz equation. 

When applying $\bm{H}_0$ along the a.s.b.\ axis, we obtain the resonance frequencies shown in Fig.~\ref{fig:setup}(e). Here, we observe a more complex evolution of the resonance frequencies for two reasons. First, for $\mu_0 H_0\lessapprox \SI{0.4}{\tesla}$, the equilibrium net magnetization is titled away from $\bm{H}_0$ and varies with $H_0$. Second, and crucially, $f_\uparrow$ and $f_\downarrow$ exhibit a pronounced avoided crossing. The dashed vertical line indicates the value of $H_0$ of minimal $\Delta f_\mathrm{res}$ (c.f. Fig.~\ref{fig:setup}(e)). 

We plot $\Delta f_\mathrm{res}$ and the half-width-at-half-maximum (HWHM) linewidths $\kappa_\uparrow$ and $\kappa_\downarrow$ as a function of the magnetic field $H_0$ in Figs.~\ref{fig:setup}(f) and (g) for the e.a.s.\ and a.s.b.\ cases, respectively. We find the mutual coupling strength $g_\mathrm{c}/2\pi=\min{|\Delta f_\mathrm{res}/2|}=\SI{0.92}{\giga\hertz}$ for the e.a.s.\ case and $g_\mathrm{c}/2\pi=\SI{6.38}{\giga\hertz}$ for the a.s.b.\ configuration. In the former case, $g_\mathrm{c}\lesssim\kappa_\uparrow,\kappa_\downarrow$ (c.f. Fig.~\ref{fig:setup}(f)). Thus, the system is in the weak to intermediate coupling regime. For the a.s.b.\ case, the linewidths $\kappa$ are at least three times smaller than $g_\mathrm{c}$. Hence the condition for strong coupling $g_\mathrm{c}>\kappa_\uparrow,\kappa_\downarrow$ is clearly satisfied. Furthermore, the extracted coupling rate of $g_\mathrm{c}/2\pi=\SI{6.38}{\giga\hertz}$ is comparable to the intrinsic excitation frequency $f_\mathrm{r}=(f_1+f_2)/2=\SI{17.2}{\giga\hertz}$. The normalized coupling rate $\eta=g_\mathrm{c}/(2\pi f_\mathrm{r})$~\cite{Niemczyk2010, Zhang2014} evaluates to $\eta=0.37$. Consequently, we observe magnon-magnon hybridization in the ultrastrong coupling regime~\cite{FriskKockum2019}. Importantly, the measured $g_\mathrm{c}$ is the intrinsic coupling strength between spin-up and spin-down magnons and is independent of geometrical factors, in particular, sample volume or filling factor. This is in stark contrast to the magnon-photon or cavity-mediated magnon-magnon coupling typically observed in spin cavitronics~\cite{Zhang2014, Maier-Flaig2017, Tavis1968, Eichler2017, ZareRameshti2018, Johansen2018}. 

\begin{figure}
	\begin{center}
		\includegraphics[]{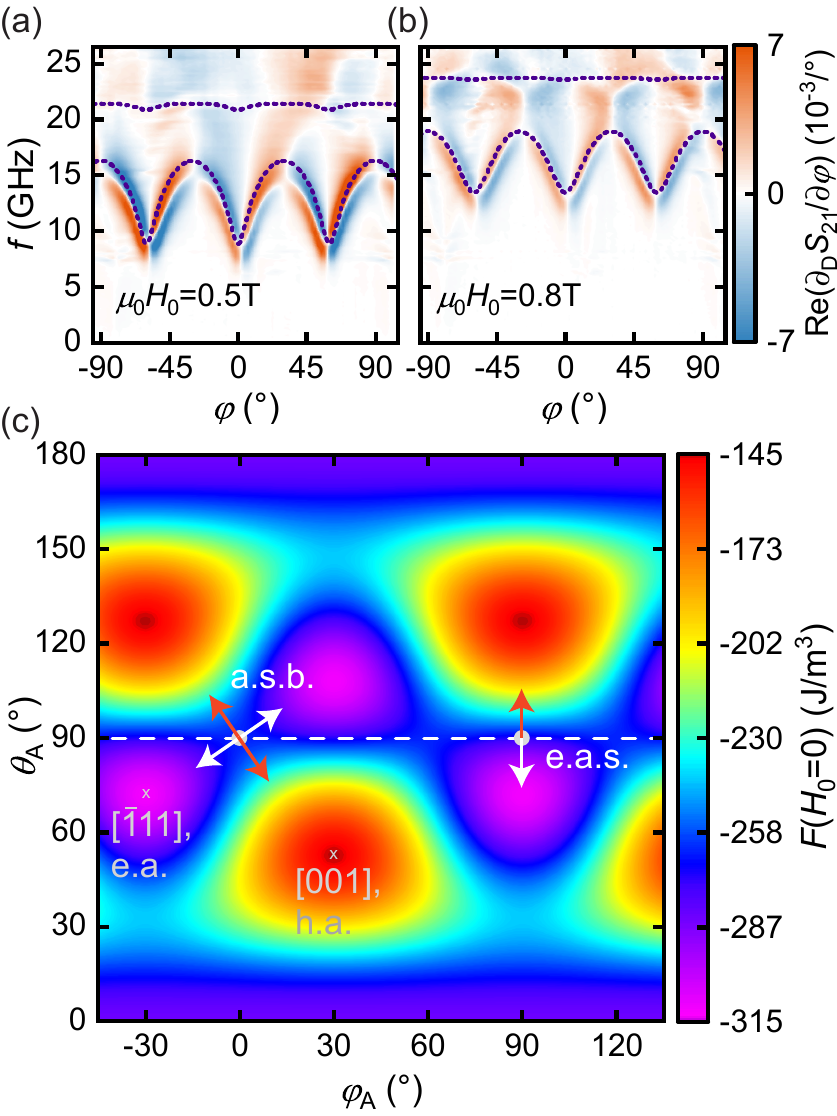}
		\caption{Tunable coupling strength and anisotropy landscape. 
		(a),(b)~BMR-data obtained with fixed magnetic field magnitudes with (a) $\mu_0H_0=\SI{0.5}{\tesla}$ (below the hybridization point) and (b) $\mu_0H_0=\SI{0.8}{\tesla}$ (above the hybridization point) as a function of the $\bm{H}_0$-orientation $\varphi$ in the (111)-disk plane at $T=\SI{280}{\kelvin}$. The blue dashed lines are the results from the numerical simulation. 
		(c)~Colormap of the free energy density $F$ for $H_0=0$. The angles $\varphi_\mathrm{A}$ and $\theta_\mathrm{A}$ denote the orientation of $M_\mathrm{A}$, defined analogously to $\varphi$ and $\theta$ in Fig.~\ref{fig:setup}(a). The dashed horizontal line at $\theta_\mathrm{A}=\ang{90}$ corresponds to the (111)-disk plane. The orange and white arrows at the e.a.s.\ ($\varphi_\mathrm{A}=\ang{90}$) and a.s.b.\ ($\varphi_\mathrm{A}=\ang{0}$) orientations point towards increasing and decreasing free energy density, respectively. The [001]-direction denotes a crystalline hard axis (h.a.) and $[\bar{1}11]$ a crystalline easy axis (e.a.).}
		\label{fig:angledep}
	\end{center}
\end{figure}
To demonstrate that the coupling is continuously tunable between the extreme cases discussed so far, we rotated $\bm{H}_0$ with fixed magnitude in the (111)-plane at $T=\SI{280}{\kelvin}$. The background corrected transmission parameter (see Supplemental Material~\cite{SupInf}) as a function of the angle~$\varphi$ is shown in Fig.~\ref{fig:angledep}(a) and (b) for $\mu_0H_0=\SI{0.5}{\tesla}$ and $\mu_0H_0=\SI{0.8}{\tesla}$, respectively. These magnetic field magnitudes correspond to $H_0$ slightly below and above the hybridization point at $T=\SI{280}{\kelvin}$ (see Fig.~S2~\cite{SupInf}). For both $H_0$ values, we observe two resonances for each value of $\varphi$, where the lower resonance frequency depends strongly on $\varphi$ while the upper one is nearly independent of $\varphi$. Overall, these results strongly indicate a $\varphi$-dependent level repulsion that allows to continuously adjust the coupling strength. 

To understand the coupling strength variation with $\varphi$, we analyze the cubic anisotropy landscape of our GdIG disk by plotting its magnetic free energy density $F$ (c.f. Eq.~(S9)~\cite{SupInf}) in Fig.~\ref{fig:angledep}(c). The applied field directions for the e.a.s.\ and a.s.b.\ cases are indicated by the two grey dots in Fig.~\ref{fig:angledep}(c). The sublattice magnetizations as well as the magnon spin axis are collinear with the applied field in our considerations. As derived rigorously below, coupling between the opposite-spin magnons is proportional to the degree of anisotropy in the free energy about the magnon spin axis~\cite{KamraPRB2017}. Moreover, since they represent small and symmetric deviations of magnetization about the equilibrium configuration, the magnons can only sense anisotropy variations that are local and averaged over antiparallel directions. Considering the a.s.b.\ configuration first, if the magnetization deviates from equilibrium along the orange (white) arrows, it experiences an increase (a decrease) in energy. Therefore, the free energy change depends on the direction of deviation and the symmetry about the magnon spin axis in this configuration is clearly broken by anisotropy. This causes a non-zero mode-coupling in the a.s.b.\ configuration. In contrast, for the e.a.s.\ configuration, an averaging over the two antiparallel directions results in a nearly vanishing and direction-independent change in the free energy, thereby effectively maintaining axial symmetry. This is prominently seen when considering the direction collinear with the orange and white arrows, which nearly cancel each other's effect on averaging. This configuration is thus named effectively axially symmetric (e.a.s.). The corresponding degree of axial anisotropy, and thus mode-coupling, varies smoothly with $\varphi$ between these two extreme cases.

The two key ingredients in the physics observed herein are (i) nonzero mode-coupling arising from violation of spin conservation by an axial anisotropy~\cite{KamraPRB2017}, and (ii) a strong amplification of the otherwise weak coupling via an exchange-enhancement effect characteristic of (quasi-)antiferromagnetic magnons~\cite{Kamra2019}. We now present a minimalistic, analytically solvable model that brings out both these pillars underlying our experiments, and yields results in good agreement with our data (Fig.~\ref{fig:setup}(d) and (e)). To this end, we employ a two-sublattice model, which corresponds to the net Fe- and Gd-sublattice in GdIG, within the Landau-Lifshitz framework and macrospin approximation, treating anisotropies as uniaxial to enable an analytical solution. In our experiments, both of the distinct anisotropy contributions considered here are provided by the cubic crystalline anisotropy of the material. Parameterizing the intersublattice antiferromagnetic exchange by $J~(> 0)$ and uniaxial anisotropies by $K~(> 0)$ and $K_\mathrm{a}$, the free energy density $F_\mathrm{m}$ is expressed in terms of the sublattice A and B magnetizations $\bm{M}_\mathrm{A,B}$, assumed spatially uniform, as
\begin{align}\label{eq:free}
F_\mathrm{m} = & - \mu_0 H_0 (M_{\mathrm{A}z} + M_{\mathrm{B}z}) \mp K \left( M_{\mathrm{A}z}^2 + M_{\mathrm{B}z}^2 \right) + K_\mathrm{a} \left( M_{\mathrm{A}x}^2 + M_{\mathrm{B}x}^2 \right) + J \bm{M}_\mathrm{A}\cdot\bm{M}_\mathrm{B},
\end{align}
where the first term is the Zeeman contribution due to the applied field $H_0 \hat{\bm{z}}$. We further assume an appropriate hierarchy of interactions $J \gg K \gg |K_\mathrm{a}|$, such that $K_\mathrm{a}$ terms do not influence the equilibrium configurations. The upper and lower signs in Eq.~\eqref{eq:free} above represent the cases of an applied field along easy and hard axes, respectively. The e.a.s.\ (a.s.b.) direction is magnetically easy (hard)~\cite{SupInf}. The axial symmetry is broken by the term proportional to $K_\mathrm{a}$, with $K_\mathrm{a}\approx 0$ for the e.a.s.\ case and $K_\mathrm{a}\neq 0$ to the a.s.b.\ case. We have choosen coordinate systems for treating the two configurations with the $z$-direction always along the applied field. The equilibrium configuration is obtained by minimizing Eq.~\eqref{eq:free} with respect to the sublattice magnetization directions (see Supplemental Material~\cite{SupInf}). The dynamics are captured by the Landau-Lifshitz equations for the two sublattices:
\begin{align}\label{eq:LL}
\frac{\partial\bm{M}_\mathrm{A,B}}{\partial t} = & - |\gamma_\mathrm{A,B}| \left[ \bm{M}_\mathrm{A,B} \times \left( - \frac{\partial F_\mathrm{m}}{\partial \bm{M}_\mathrm{A,B}} \right) \right],
\end{align}
where $\gamma_\mathrm{A,B}$ are the respective sublattice gyromagnetic ratios, assumed negative. It is convenient to employ a new primed coordinate system with equilibrium magnetizations collinear with $\hat{\bm{z}}^\prime$. The ensuing dynamical equations are linearized about the equilibrium configuration which, on switching to Fourier space (i.e. $M_{\mathrm{A}x^\prime} = m_{\mathrm{A}x^\prime} e^{i \omega t}$ and so on), lead to the coupled equations describing the eigenmodes expressed succinctly as a 4$\times$4 matrix equation:
\begin{align}
\left(\tilde{P}_0 + \tilde{P}_\mathrm{a} \right) \tilde{m} = & 0,
\label{eq.p_matrices}
\end{align}
where $\tilde{m}^\intercal = [m_{\mathrm{A}+}~ m_{\mathrm{B}+} ~ m_{\mathrm{A}-} ~ m_{\mathrm{B}-}] $ with $m_{\mathrm{A}\pm} \equiv m_{\mathrm{A}x^\prime} \pm i m_{\mathrm{A}y^\prime}$ and so on. The matrix $\tilde{P}_0$ is block diagonal in $2\times2$ sub-matrices and describes the uncoupled spin-up and spin-down modes, distributed over both sublattices. The matrix $\tilde{P}_\mathrm{a}$ captures axial-symmetry-breaking anisotropy effects, and provides the spin-nonconserving, off-diagonal terms that couple the two modes and underlie the hybridization physics at play. The detailed expressions for the matrices are provided in the Supplemental Material~\cite{SupInf}.

For applied fields along the easy-axis (e.a.s.), the equilibrium configuration is given by $\bm{M}_\mathrm{A} = M_{\mathrm{A}0} \hat{\bm{z}}$ and $\bm{M}_\mathrm{B} = - M_{\mathrm{B}0} \hat{\bm{z}}$, with $M_{\mathrm{A}0,\mathrm{B}0}$ the respective sublattice saturation magnetizations and $M_{\mathrm{A}0} \gtrsim M_{\mathrm{B}0}$.  
For the case of a sufficiently small field applied along the hard axis (a.s.b.), the equilibrium orientation of $\bm{M}_\mathrm{A}$ is orthogonal to the hard axis. With increasing field strength, $\bm{M}_\mathrm{A}$ moves to align with the applied field. In the considered temperature and field range, $\bm{M}_\mathrm{B}$ always remains essentially anticollinear to $\bm{M}_\mathrm{A}$~\cite{Ganzhorn2016}. The initial decrease of the resonance mode with lower frequency (Fig.~\ref{fig:setup}(e)) is associated with this evolution of the equilibrium configuration. The frequency dip signifies alignment of equilibrium $\bm{M}_\mathrm{A}$ with the $z$-axis. 
Only the $K_\mathrm{a}$ anisotropy term breaks axial symmetry about the equilibrium magnetization direction ($z$-axis) and leads to off-diagonal terms in $\tilde{P}_\mathrm{a}$, which couples the two modes. The coupling-mediated frequency splitting $\Delta f_\mathrm{res}$, where uncoupled eigenmode frequencies would cross, is evaluated employing Eq.~\eqref{eq.p_matrices} as:
\begin{align}\label{eq:delf}
2 \pi \Delta f_\mathrm{res} =  & \omega_\mathrm{c} ~ \sqrt{\frac{16 J M_0^2}{J \left( M_{\mathrm{A}0} - M_{\mathrm{B}0} \right)^2 + F_{\mathrm{eq}}  }},
\end{align}
where $\omega_\mathrm{c} \equiv |\gamma| |K_\mathrm{a}| M_0$ is the bare coupling rate, considering $\gamma_\mathrm{A} \approx \gamma_\mathrm{B} \equiv \gamma$ and $M_{\mathrm{A}0} \approx M_{\mathrm{B}0} \equiv M_0$ near the compensation point. $F_{\mathrm{eq}}$, given by $16 K M_0^2$ for $H_0$ along an easy axis, is an equivalent free energy density comparable to the anisotropy contribution, parametrized by $K$. The bare coupling rate is thus enhanced by a maximum value of $\sqrt{J/K}$ at the compensation point yielding a greatly enlarged coupling. Hereby a small coupling of $\omega_\mathrm{c}=2\pi\cdot\SI{160}{\mega\hertz}$ originating from a weak cubic anisotropy present in GdIG is greatly enhanced as demonstrated by Eq.~(\ref{eq:delf}) and the analytical model results displayed in Fig.~\ref{fig:setup}(e), quantitatively describing our experimental observations. The amplification of coupling from 160 MHz to several GHz is an exchange-enhancement effect~\cite{Rodrigue1960, Keffer1952, Kamra2018, Kamra2019}. This (exchange-)enhancement is an embodiment of antiferromagnetic quantum fluctuations~\cite{Kamra2019} predicted similarly to amplify magnon-mediated superconductivity~\cite{Erlandsen2019}. 

Our findings demonstrate that previously typically neglected details of the magnetocrystalline anisotropy can lead to giant effects on spin-dynamics if they have the appropriate symmetry and are exchange-enhanced. The ultrastrong and size-independent magnon-magnon coupling reported here opens exciting perspectives for studying ultrastrong coupling effects in nanoscale devices and exploring quantum-mechanical coupling phenomena beyond classical electrodynamics. The reported effect also enables the tuning and tailoring of quasi-antiferromagnetic dynamics and magnons. 

Note added: During the preparation of the manuscript, we became aware of a related study showing magnon-magnon coupling in the canted antiferromagnet CrCl$_3$~\cite{MacNeill2019}.  

{\it Acknowledgments.} -- We thank A. Habel, K. Helm-Knapp, and K. Danielewicz for technical support. We gratefully acknowledge the financial support of the Deutsche Forschungsgemeinschaft (DFG, German Research Foundation) via Germany's Excellence Strategy – EXC-2111–-390814868 (R.G. and H.H.) and the projects WE5386/4 and WE5386/5 (L.L. and M.W.). A.K. acknowledges financial support from the Research Council of Norway through its Centers of Excellence funding scheme, project 262633, ``QuSpin''. W.B. was supported by the DFG through SFB 767 and thanks the Center of Excellence „QuSpin“ by the Research Council of Norway and Arne Brataas (NTNU Trondheim) for hospitality.

\end{document}